\documentclass[conference,letterpaper]{IEEEtran}

\usepackage[utf8]{inputenc} 
\usepackage[T1]{fontenc}
\usepackage{url}
\usepackage{ifthen}
\usepackage{cite}
\usepackage[cmex10]{amsmath} 
\usepackage{amsfonts,amssymb}
\usepackage{graphicx}
\usepackage{color}
\usepackage{flushend}

\ifCLASSOPTIONcompsoc
 \usepackage[caption=false,font=normalsize,labelfont=sf,textfont=sf]{subfig}
\else
 \usepackage[caption=false,font=footnotesize]{subfig}
\fi

\newtheorem{theorem}{Theorem}
\newtheorem{lemma}{Lemma}

\newtheorem{remark}{Remark}
\newtheorem{construction}{Construction}

\newtheorem{example}{Example}
\newtheorem{definition}{Definition}
\newtheorem{proposition}{Proposition}

\newcommand{\ie}{{\it i.e.}}
\newcommand{\eg}{{\it e.g.}}

\newcommand{\code}{\mathcal{C}}

\newcommand{\R}{\mathbb{R}} 
\newcommand{\Rd}{\mathbb{R}^d} 

\newcommand{\loss}[1]{\ell\left(#1\right)} 
\newcommand{\w}{\mathbf{w}} 
\newcommand{\xsubi}{\mathbf{x}_i} 
\newcommand{\ysubi}{y_i} 

\newcommand{\xj}[1]{\mathbf{x}_{#1}} 
\newcommand{\yj}[1]{y_{#1}} 

\newcommand{\watt}[1]{\mathbf{w}^{(#1)}} 
\newcommand{\gi}[1]{\boldsymbol{g}_{#1}} 
\newcommand{\Di}[1]{D_{#1}} 
\newcommand{\Wi}[1]{W_{#1}} 

\newcommand{\norm}[1]{\left\lVert#1\right\rVert_2^2} 

\newcommand{\mat}[1]{\boldsymbol{#1}} 
\newcommand{\I}[1]{\boldsymbol{I}_{#1}} 
\newcommand{\Jc}[1]{\boldsymbol{1}_{#1}} 
\newcommand{\J}[2]{\boldsymbol{J}_{#1 \times #2}} 
\newcommand{\Js}[1]{\boldsymbol{J}_{#1}} 
\newcommand{\Ze}[2]{\boldsymbol{0}_{#1 \times #2}} 
\newcommand{\Zs}[1]{\boldsymbol{0}_{#1}} 

\newcommand{\set}[1]{\mathcal{#1}} 

\newcommand{\supp}[1]{\mathrm{supp}\left(#1\right)} 
\newcommand{\errF}[1]{\mathrm{err}_{\set{F}}\left(#1\right)} 
\newcommand{\errs}[1]{\mathrm{err}_{\sgc}\left(#1\right)} 
\newcommand{\vopt}{\mat{v}_{\mathrm{opt}}} 

\newcommand{\bibd}{(\vd,\bd,\kd,\rd,\lamd)} 
\newcommand{\vd}{v} 
\newcommand{\bd}{b} 
\newcommand{\kd}{k} 
\newcommand{\rd}{r} 
\newcommand{\lamd}{\lambda} 

\newcommand{\ngc}{N} 
\newcommand{\kgc}{K} 
\newcommand{\lgc}{L} 
\newcommand{\rgc}{R} 
\newcommand{\sgc}{S} 
\newcommand{\nsgc}{\bar{S}} 
\newcommand{\sgt}[1]{S^{*}(#1)} 
\newcommand{\sgtLB}[1]{S^{*}_{LB}(#1)} 

\newcommand{\matEF}{\mat{E}_{\set{F}}} 
\newcommand{\Jh}{\mat{\hat{J}}} 
\newcommand{\Jhs}[1]{\mat{\hat{J}}_{#1}} 

\newcommand{\Neb}[1]{\set{N}\left(#1\right)} 

\newcommand\blfootnote[1]{%
  \begingroup
  \renewcommand\thefootnote{}\footnote{#1}%
  \addtocounter{footnote}{-1}%
  \endgroup
}

\begin{document}
\title{Gradient Coding Based on Block Designs for Mitigating Adversarial Stragglers} 

 \author{%
   \IEEEauthorblockN{Swanand Kadhe, O. Ozan Koyluoglu, and Kannan Ramchandran}
   \IEEEauthorblockA{Department of Electrical Engineering and Computer Sciences,\\ 
   				University of California, Berkeley\\
                     Emails: \{swanand.kadhe, ozan.koyluoglu, kannanr\}@berkeley.edu}
 }

\maketitle

\begin{abstract}
Distributed implementations of gradient-based methods, wherein a server distributes gradient computations across worker machines, suffer from slow running machines, called \emph{stragglers}. Gradient coding is a coding-theoretic framework to mitigate stragglers by enabling the server to recover the gradient sum in the presence of stragglers. \emph{Approximate gradient codes} are variants of gradient codes that reduce computation and storage overhead per worker by allowing the server to approximately reconstruct the gradient sum. 

In this work, our goal is to construct approximate gradient codes that are resilient to stragglers selected by a computationally unbounded adversary. {Our motivation for constructing codes to mitigate adversarial stragglers stems from the challenge of tackling stragglers in massive-scale elastic and serverless systems, wherein it is difficult to statistically model stragglers.}
Towards this end, we propose a class of approximate gradient codes based on balanced incomplete block designs (BIBDs). We show that the \emph{approximation error} for these codes depends only on the number of stragglers, and thus, adversarial straggler selection has no advantage over random selection. 
In addition, the proposed codes admit computationally efficient decoding at the server. Next, to characterize fundamental limits of adversarial straggling, we consider the notion of \emph{adversarial threshold} -- the smallest number of workers that an adversary must straggle to inflict certain approximation error. We compute a lower bound on the adversarial threshold, and show that codes based on symmetric BIBDs maximize this lower bound among a wide class of codes, making them excellent candidates for mitigating adversarial stragglers.
\end{abstract}


\blfootnote{This work is supported in part by National Science Foundation grants CCF-1748585 and CNS-1748692. 
}

\section{Introduction}
\label{sec:intro}







In many real-world applications, the size of training datasets has grown significantly over the years to the point that it is becoming crucial to implement learning algorithms in a distributed fashion. 
However, in practice the gains due to parallelization are often limited due to {\it stragglers} -- workers that are slowed down due to unpredictable factors such as network latency, hardware failures, etc.~\cite{Hoefler:10:noise,Dean:13:tail}. 
For instance, recent studies~\cite{Tandon:17,Yadwadkar:16} have demonstrated that straggling machines may run $\times 5$ to $\times 8$ slower than a typical machine on Amazon EC2. The straggler problem is even more daunting in massive-scale computing systems such as~\cite{Jonas:17}, which use AWS Lambda. 
Left untreated, stragglers severely impact latency, as the performance in each iteration is determined by the slowest machine. 

Conventional approaches to mitigate stragglers involve detecting stragglers, ignoring stragglers, or replicating jobs across workers  (see, \eg,~\cite{Ananthanarayanan:10:matri,Ananthnarayanan:13:clones,Wang:Joshi:15,Chen:16}). 
Recently, using coding-theoretic ideas to mitigate stragglers has gained significant research attention, see, \eg,~\cite{Lee:18,Dutta:16,Avestimehr:18:coded-matrix,Aktas:17} for distributed computing, and~\cite{Tandon:17,Halbawi:17,Raviv:18,YeAbbe:18i,Avestimehr:17} for distributed learning. 

A coding theoretic framework for mitigating stragglers in distributed gradient-based learning methods was first proposed in~\cite{Tandon:17}. The setup consists of $\ngc$ worker machines and a parameter server. Training examples are partitioned into $\kgc$ parts, and every worker is assigned $\lgc$ of the $\kgc$ parts. Each worker computes the partial gradient on its assigned examples, linearly combines the results according to some pre-specified vector of coefficients, and returns the result to the server. Note that the parameter $\lgc$ essentially specifies the computation and storage load on individual workers. The authors  showed that by redundantly assigning the parts across the workers and by judiciously choosing the coefficients of the linear combination at each worker, it is possible to {\it exactly} recover the sum of all gradients even if any $\sgc$ workers straggle, and fail to return their results. 
Alternate code constructions for gradient coding have been proposed in~\cite{Dutta:16,Halbawi:17,Raviv:18,YeAbbe:18}.

Gradient coding schemes designed for exactly recovering the gradient sum have two limitations. First, they fundamentally require heavy computational and storage overhead at each worker. In particular, in~\cite{Tandon:17}, it was established that any coding scheme designed to tolerate $\sgc$ stragglers must have $\lgc \geq \kgc(\sgc + 1)/\ngc$. This implies that the higher the straggler tolerance required, the larger is the computation and storage overhead per worker. Second, since the schemes are designed for a particular number of stragglers $\sgc$, it is necessary to have an estimate on $\sgc$ at the design time. This is not feasible for many practical schemes as straggler behavior can vary unpredictably.

In~\cite{Raviv:18}, the authors showed that these limitations can be lifted by allowing the server to approximately recover the gradient sum.
Indeed, in many practical learning algorithms, it is sufficient to approximately reconstruct the gradient sum. 
The authors construct codes based on expander graphs, for which, the $\ell_2$-error of the approximate gradient sum, referred to as {\it approximation error}, degrades gracefully with the number of stragglers.
These so-called {\it approximate gradient codes} 
do not require to have an estimate of the number of stragglers $\sgc$ {\it a priori}, and allow the computation and storage overhead per worker to be substantially small.

In~\cite{CharlesP:17}, the authors evaluate three families of approximate gradient codes: fractional repetition codes (FRCs), Bernoulli gradient codes (BCGs), and regularized BCGs based on sparse random graphs. They show that FRCs achieve small approximation error when the stragglers are chosen at random. However, FRCs perform poorly for adversarial stragglers, wherein an adversary can force to straggle any subset of workers up to a given size. Further, it is shown that adversarial straggler selection in general codes is NP-hard. In~\cite{CharlesP:18}, the authors propose stochastic block codes (SBCs), which make it difficult for a computationally limited (polynomial-time bounded) adversary to select stragglers.

In this work, our goal is construct approximate gradient codes that can mitigate adversarial stragglers even for a computationally unbounded adversary. {Our key idea is to construct codes based on combinatorial block designs. A block design is a family of subsets of a (finite) set, chosen in such a way that certain symmetry properties are satisfied (see~\cite{Stinson:2003} for details)}. We note that codes resilient to adversarial stragglers are useful in massive-scale elastic and serverless systems (such as~\cite{Jonas:17}), wherein it is difficult to statistically model stragglers. Furthermore, we are interested in understanding fundamental limits of adversarial straggler selection. 

{\bf Our Contributions:} We propose a class of gradient codes based on balanced incomplete block designs (BIBDs) for mitigating adversarial stragglers. We show that the approximation error for these codes depends only on the number of stragglers, and not on which specific set of workers is straggling. Therefore, an adversary that can intelligently select stragglers has no advantage over one that chooses an arbitrary subset of stragglers. Moreover, for the proposed codes, the {\it decoding vector} at the server can be computed in closed-form. This enables the server to perform the decoding in a computationally efficient manner.

Next, we define the notion of {\it adversarial threshold} for a gradient code. The adversarial threshold specifies the minimum number of workers that an adversary must straggle to enforce that the approximation error is above a given target. 
{We compute a lower bound on the adversarial threshold. Further, we show that codes based on symmetric BIBDs are excellent candidates for mitigating adversarial stragglers, since they maximize this lower bound among a wide class of codes.}

\section{Framework}
\label{sec:framework}

{\bf Notation:} We use standard script for scalars, bold script for vectors and matrices, and calligraphic letters for sets. For a positive integer $n$, let $[n] = \{1,2,\ldots,n\}$. 
For a vector $\mat{v}$, let $\supp{\mat{v}}$ denote the support of $\mat{v}$. For a matrix $\mat{H}$, let $\mat{H}^T$ be its transpose, $\mat{H}^{\dagger}$ be its Moore-Penrose inverse, $\mat{H}_{i,j}$ be its $(i,j)$-th entry, $\mat{H}_j$ be its $j$-th column, and $\mat{H}_{\set{T}}$ be the sub-matrix of $\mat{H}$ corresponding to the columns indexed by a set $\set{T}\subset[n]$. 
Let $\Jc{m}$ denote the $m\times 1$ all ones vector, and $\I{m}$ denote the $m\times m$ identity matrix. Let $\J{m}{n}$ and $\Ze{m}{n}$ denote the $m\times n$ all ones and all zero matrices, respectively; when $m = n$, we simplify the  notation to $\Js{m}$ and $\Zs{m}$. 

\subsection{Distributed Training}
\label{sec:training}
The process of learning the parameters $\w\in\Rd$ of a model given a dataset $D = \{(\xsubi,\ysubi)\}_{i=1}^{M}$ of $M$ samples, where $\xsubi \in \Rd$ and $\ysubi\in \R$, can be cast as the {\it empirical risk minimization} (ERM) problem given as 
 \begin{equation}
 \label{eq:objective}
 \min_{\w} \frac{1}{M}\sum_{i=1}^{M} \loss{\xsubi,\ysubi;\w},
 \end{equation}
where $\loss{\xsubi,\ysubi;\w}$ is a loss function that measures the accuracy of the prediction made by the model $\w$ on the sample $(\xsubi,\ysubi)$. 

One popular method to approximately solve the ERM is stochastic gradient descent (SGD). SGD begins with some initial guess of $\w$ as $\w^{(0)}$, and then iteratively updates the parameters as $\watt{t+1} = \watt{t} - \alpha_t \nabla\loss{\xj{i_t},\yj{i_t};\watt{t}}$, where $i_t$ is a  sample index chosen randomly from $[M]$, and  $\alpha_t > 0$ is the learning rate (or step size) at iteration $t$.

In a distributed setting, it is possible to take advantage of parallelism by using mini-batch SGD. In every iteration of mini-batch SGD, a (possibly random) subset $S_t$ of $B$ samples is chosen and the model is updated as \mbox{$\watt{t+1} = \watt{t} - \frac{\alpha_t}{B}\sum_{i\in S_t} \nabla\loss{\xsubi,\ysubi;\watt{t}}$.}

Next, we describe the framework of gradient coding which mitigates stragglers in a distributed implementation of mini-batch SGD by redundantly assigning gradients to workers.

 \begin{remark}
 \label{rem:SGD-vs-others}
 Even though we focus on mini-batch SGD for the ease of exposition, our proposed coding techniques can be applied to other common first-order methods in machine learning. Moreover, our techniques are applicable to any distributed algorithm that requires the sum of multiple functions.  
 \end{remark}

\subsection{Gradient Coding}
\label{sec:grad-coding}
Consider a distributed master-worker setting consisting of $\ngc$ worker machines $\Wi{1}$, $\Wi{2}$, $\ldots$, $\Wi{\ngc}$, and a parameter server. 
 We focus our attention to a given iteration $t$, and fix a batch of $B$ samples $S_t$. Without loss of generality, assume that $S_t = \{1,2,\ldots,B\}$. We omit the explicit dependence on the iteration $t$ hereafter, since our focus is on a given iteration. 
 
We partition the batch into $\kgc$ subsets of equal size\footnote{We assume $\kgc\mid B$ for simplicity. Our schemes can be easily adapted when $\kgc \nmid B$.}, denoted as $\Di{1},\Di{2},\ldots,\Di{\kgc}$. Define the gradient vector of the partial data $\Di{i}$, called {\it partial gradient}, as $\gi{i} := \sum_{\mathbf{x}_j,\mathbf{y}_j\in\Di{i}} \nabla \loss{\xj{j},\yj{j};\w}$. Note that the server is interested in computing $\mat{g} := \sum_{i=1}^{\kgc}\gi{i}$. 

A gradient code (GC) consists of an encoding matrix $\mat{E}\in\mathbb{R}^{\kgc\times \ngc}$. 
The $j$-th column $\mat{E}_j$ of $\mat{E}$ corresponds to worker $j$, and determines which samples are assigned to the worker and what linear combination of gradients it returns to the server. In particular, let $\supp{\mat{E}_j} = L_j$. Then, the $j$-th worker is assigned the subsets $\{\Di{i} : i\in L_j\}$, and it sends back to the server $\mat{c}_j = \frac{1}{\kgc}\sum_{i\in L_j} \gi{i}\mat{E}_{i,j}$. 

Let $\lgc = \max_{j\in[\ngc]}\left|\supp{\mat{E}_j}\right|$ and $\rgc = \min_{j\in[\kgc]}\left|\supp{(\mat{E}^T)_j}\right|$. We refer to $\lgc$ as the computation load of $\mat{E}$ since a worker works on at most $\lgc$ gradients. Note that for load balancing, it is good to assign the same number of gradient computations to each worker. 
We refer to $\rgc$ as the replication factor of $\mat{E}$ since each gradient is computed by at least $\rgc$ workers. We denote such a gradient code as an $(\ngc,\kgc,\lgc,\rgc)$-GC. (We summarize the notation in Table~\ref{tab:gradient-coding}.) 

\begin{table}[!t]
    \renewcommand{\arraystretch}{1.2}
    \caption{Notation for Gradient Coding}
    \label{tab:gradient-coding}
    \centering
    \begin{tabular}{|c|l|}
    \hline
    $N$ & Number of workers\\
    $K$ & Number of data partitions\\
    $L$ & Computational load per worker\\
    $R$ & Replication factor\\
    $\mat{E}$  & Encoding matrix of size $K\times N$\\
    \hline
    \end{tabular}
\end{table}

Decoding consists of finding a linear combination of the results from the non-straggling workers to {\it approximate} the gradient sum $\mat{g}$. Specifically, given a set of non-stragglers $\set{F}\subset[\ngc]$ of size $|\set{F}| = \ngc-\sgc$, the server finds a vector $\mat{v}\in\mathbb{R}^{\ngc-\sgc}$, and computes $\hat{\mat{g}} = \mat{C}_{\set{F}} \mat{v}$, where $\mat{C} = \begin{bmatrix} \mat{c}_1 & \mat{c}_2 & \cdots & \mat{c}_{\ngc}\end{bmatrix}$.

Next, we use the framework of~\cite{CharlesP:17} (see also~\cite{Raviv:18,CharlesP:18}) to define the approximation error and the optimal decoding vector for a given gradient code as follows. 

\begin{definition}
\label{def:decoding-error}
Given an encoding matrix $\mat{E}$, the approximation error $\errF{\mat{E}}$ for a given set of non-stragglers $\set{F}\subseteq[\ngc]$ of size $\ngc-\sgc$ is defined as
\begin{equation}
    \label{eq:errF}
    \errF{\mat{E}} = \min_{\mat{v}\in\mathbb{R}^{\ngc-\sgc}}\norm{\mat{E}_{\set{F}}\mat{v} - \Jc{\kgc}},
\end{equation}
and a solution $\vopt$ to~\eqref{eq:errF} is called an {\it optimal decoding vector}. The worst-case approximation error for $\sgc$ $(<\ngc)$ stragglers is defined  as
\begin{equation}
    \label{eq:vopt}
    \errs{\mat{E}} = \max_{\substack{\set{F}\subset[\ngc]\\ |\set{F}|=\ngc-\sgc}}\min_{\mat{v}\in\mathbb{R}^{\ngc-\sgc}}\norm{\mat{E}_{\set{F}}\mat{v} - \Jc{\kgc}}.
\end{equation}
\end{definition}

Note that the deviation of $\hat{\mat{g}}$ from $\mat{g}$ can be bounded in terms of $\errF{\mat{E}}$ as  as $\norm{\hat{\mat{g}} - \mat{g}} \leq \norm{\mat{G}}\errF{\mat{E}}$, where $\mat{G}$ is the matrix consisting of all the gradient vectors~\cite{Raviv:18}.
Our goal is to construct encoding matrices such that the worst-case approximation error is small. In addition, it is desirable if an optimal decoding vector can be computed efficiently.  



\section{Preliminaries on Block Designs}
\label{sec:designs}
We briefly review the relevant notions from the theory of block designs. For details, we refer the reader to~\cite{Stinson:2003}.

\begin{definition}
\label{def:bibd}
[{\it Design and Incidence Matrix}] 
A design is a pair $(X,\mathcal{A})$, where $X$ is a set of elements called {\it points}, and $\mathcal{A}$ is a collection of nonempty subsets of $X$ called {\it blocks}. 
Consider a design $(X,\mathcal{A})$ with $X = \{x_1,x_2,\ldots,x_{\vd}\}$ and $\mathcal{A} = \{A_1,A_2,\ldots,A_{\bd}\}$. Then, the {\it incidence matrix} of $(X,\mathcal{A})$ is a $\vd\times \bd$ binary matrix $\mat{M}$ such that $\mat{M}_{i,j} = 1$ if $x_i\in A_j$ and $\mat{M}_{i,j} = 0$ if $x_i\notin A_j$. 
\end{definition}

Balanced incomplete block designs are probably the most-studied type of designs. They are  defined as follows.
\begin{definition}
\label{def:bibd}
[{\it BIBD}] 
A $\bibd$-{\it balanced incomplete block design (BIBD)} is a design  $(X,\mathcal{A})$ with $v$ points and $b$ blocks, each of size $k$, such that 
every point is contained in exactly $\rd$ blocks and any pair of distinct points is contained in exactly $\lamd$ blocks. (We summarize the notation in Table~\ref{tab:BIBD}.)
\end{definition}

\begin{remark}
\label{rem:incidence-matrix}
Note that the incidence matrix of a $\bibd$-BIBD is such that its every column contains exactly $\kd$ ones, every row contains exactly $\rd$ ones, and any two distinct rows intersect in exactly $\lamd$ locations. 
It is well-known that the parameters $\vd$, $\bd$, $\kd$, $\rd$, and $\lamd$ of a $\bibd$-BIBD should be such that $\vd\rd = \bd\kd$ and $\rd(\kd-1) = \lamd(\vd-1)$. 
\end{remark}

\begin{table}[!t]
    \renewcommand{\arraystretch}{1.2}
    \caption{Notation for BIBDs}
    \label{tab:BIBD}
    \centering
    \begin{tabular}{|c|l|}
    \hline
    $\vd$ & Number of points\\
    $\bd$ & Number of blocks\\
    $\kd$  & Number of points per block\\
    $\rd$ & Number of blocks containing a point\\
    $\lambda$ & Number of blocks containing a pair of points\\
    $\mat{M}$ & Incidence matrix of size $\vd \times \bd$\\
    \hline
    \end{tabular}
\end{table}


\begin{example}
\label{ex:Fano}
[Fano Plane] A (symmetric) $(7,7,3,3,1)$-BIBD: $X = \{1,2,\ldots,7\}$ and $\mathcal{A} = \{123,145,167,246,257,347,356\}$. (To save space, we write blocks in the form $abc$ rather than $\{a,b,c\}$.) Observe that every block contains 3 points, and every point occurs in 3 blocks. In addition, every pair of distinct points is contained in exactly one block.
\end{example}






\section{Gradient Codes Using BIBDs}
\label{sec:code-bibd}
In this section, we consider gradient codes based on BIBDs. For any $\bibd$-BIBD $(X = \{X_1,\ldots,X_{\vd}\}, \set{A} = \{\set{A}_1,\ldots,\set{A}_{\bd}\})$, let us construct a gradient code using the BIBD in the following natural way. Consider a distributed system with $\ngc = \bd$ workers. Partition the training dataset into $\kgc = \vd$ subsets $\Di{1},\Di{2},\ldots,\Di{\vd}$, and allocate a subset $\Di{i}$ to worker $j$ if the $i$-th point belongs to the $j$-th block, \ie, if $X_i\in\set{A}_j$. 
By the definition of a BIBD, each worker will compute $\lgc = \kd$ gradients and each gradient will be computed $\rgc = \rd$ times. This construction can be concisely described in terms of the incidence matrix as follows.

\begin{construction}
\label{con:BIBD-code}
Given a $\bibd$-BIBD with incidence matrix $\mat{M}$, construct a gradient code with the encoding matrix  $\mat{E} = \mat{M}$. 
The resulting gradient code is an $(\ngc = \bd, \kgc = \vd, \lgc = \kd, \rgc = \rd)$-GC.
\end{construction}

{We note that the parameters of any code constructed using a BIBD are restricted to $\ngc\lgc = \kgc\rgc$ and $\rgc(\lgc-1) = \lamd(\kgc-1)$ (see Remark~\ref{rem:incidence-matrix}). On the other hand, since BIBDs have received significant research attention in combinatorics and a large number of constructions have been proposed (see \eg~\cite{Handbook:06}), this enables us to construct a class of gradient codes for a wide range of parameters. 

In the following sections, we focus our attention to codes constructed using three well-studied families of BIBDs. We show that these codes have two key advantages. First,  combinatorial properties of BIBDs enable us to compute an optimal decoding vector in closed-form. This results in extremely simple and efficient decoding at the server. Secondly, these codes are resilient to adversarial stragglers. 
}


\subsection{Gradient Codes Using Symmetric BIBDs}
\label{sec:symmetric-BIBDs}
Symmetric designs form an important class of block designs. The well-known Fisher's inequality for BIBDs states that, for any $\bibd$-BIBD, the parameters $\vd$ and $\bd$ should satisfy $\bd \geq \vd$. 
A $\bibd$-BIBD in which $\vd = \bd$ (or, equivalently $\rd = \kd$) is called a {\it symmetric} $\bibd$-BIBD.\footnote{It is worth noting that the incidence matrix of a symmetric BIBD need not be a symmetric matrix.} 
Using symmetric BIBDs in Construction~\ref{con:BIBD-code} results in a class of gradient codes with $\ngc = \kgc = \vd$,\footnote{We note that the parameter regime $\ngc = \kgc$ has received primary research attention, see \eg~\cite{Tandon:17,Raviv:18,CharlesP:17,CharlesP:18,YeAbbe:18i}.} and $\lgc = \rgc = \kd$. 

 \begin{remark}
 \label{rem:BIBD-code}
It is well-known that any pair of distinct blocks of a symmetric $\bibd$-BIBD intersect in exactly $\lambda$ points. This ensures that for a gradient code constructed using a symmetric BIBD, any pair workers share exactly $\lambda$ gradients. 
 \end{remark}
This property enables us to characterize the approximation error as well as optimal decoding vector in closed form.




\begin{theorem}
\label{thm:BIBD-code}
Consider an $(\ngc,\kgc,\lgc,\rgc)$-GC 
obtained from a symmetric BIBD using Construction~\ref{con:BIBD-code}. 
\begin{enumerate}
    \item For any set of non-stragglers $\set{F}$ of size $(\ngc - \sgc)$, an optimal decoding vector is 
\begin{equation}
    \label{eq:vopt-BIBD-code}
    \vopt = \left(\frac{\lgc}{\lgc + \lamd(\ngc - \sgc - 1)}\right)\Jc{\ngc-\sgc}.
\end{equation}
    \item The worst-case approximation error for $\sgc$ stragglers is 
\begin{equation}
    \label{eq:errs-BIBD-code}
    \errs{\mat{E}} = \kgc - \frac{\lgc^2(\ngc - \sgc)}{\lgc + \lamd(\ngc - \sgc - 1)}.
\end{equation}
\end{enumerate} 
\end{theorem}
\begin{IEEEproof}
(Sketch) 
The key idea is to show that for any set of non-stragglers of given size, it is possible to solve the normal equation. This relies on the property that any pair of workers share exactly $\lamd$ gradient computations. The main technical tool that we use is a {\it matrix inversion lemma} from~\cite{Miller:81:matrix-inversion} for the inverse of the sum of two matrices (see Appendix~\ref{app:matrix-inversion-lemma}). The complete proof is deferred to Appendix~\ref{app:BIBD-code}.
\end{IEEEproof}

\begin{remark}
\label{rem:symmetric-BIBD-code-adversarial}
{
Note that the optimal decoding vector~\eqref{eq:vopt-BIBD-code} and worst-case decoding error~\eqref{eq:errs-BIBD-code} depend only on the number of stragglers and not on the specific set of stragglers. Therefore, decoding at the server can be performed in a very efficient way. Moreover, since any set of $\sgc$ stragglers is as harmful as other (in terms of the approximation error), an adversary cannot do better than straggling an arbitrary set of $S$ stragglers. This makes these codes resilient to adversarial straggling.
}
\end{remark}

Note that it is possible to construct several classes of gradient codes based on well-known families of symmetric BIBDs. We present a few examples in the following.

\subsubsection{Class of Gradient Codes Based on Projective Geometries} For any prime power $q$ and integer $m\geq 2$, the projective geometry of order $q$ and dimension $m$ can be used to obtain a symmetric BIBD~\cite{Stinson:2003}. Using such a BIBD in Construction~\ref{con:BIBD-code} yields a class of gradient codes with the following parameters: $\ngc = \kgc = (q^{m+1} - 1)/(q-1)$, $\lgc = \kgc = (q^m - 1)/(q - 1)$. Any pair of distinct workers share exactly $\lamd = (q^{m-1} - 1)/(q - 1)$ gradient computations. 

As we will see in Sec.~\ref{sec:numerical}, gradient codes based on projective planes (\ie, $m = 2$) are nearly optimal in terms of the worst-case approximation error when $\sgc = O(q)$ and $q$ is sufficiently large. 

\subsubsection{Class of Gradient Codes Based on Hadamard Designs} For a positive integer $m \geq 2$, a symmetric BIBD can be constructed using a Hadamard matrix of order $4m$~\cite{Stinson:2003}. Using such a BIBD in Construction~\ref{con:BIBD-code} yields a class of gradient codes with the following parameters: $\ngc = \kgc = 4m - 1$, $\lgc = \kgc = 2m - 1$. Any pair of distinct workers share exactly $\lamd = m - 1$ gradient computations. 

\subsection{Gradient Codes Using Dual Designs}
\label{sec:dual-BIBDs}



Given a $\bibd$-BIBD $(X,\set{A})$ with the incidence matrix $\mat{M}$, the design having incidence matrix $\mat{M}^T$ is called the {\it dual design} of $(X,\set{A})$. When the dual of a $\bibd$-BIBD is used in Construction~\ref{con:BIBD-code}, it is easy to see that the resulting code is a $(\ngc = \vd, \kgc = \bd,\lgc = \rd, \rgc = \kd)$-GC. Note that, unlike symmetric BIBDs, using dual designs allows us to construct codes for which $\ngc \ne \kgc$.

Since every pair of distinct points is contained in $\lamd$ number of blocks in a BIBD, 
any two distinct blocks of the dual intersect in exactly $\lamd$ points. 

\begin{theorem}
\label{thm:dual-BIBD-code}
Consider an $(\ngc,\kgc,\lgc,\rgc)$-GC with encoding matrix $\mat{E}$ obtained from the dual of a BIBD using Construction~\ref{con:BIBD-code}. Then, for any set of non-stragglers $\set{F}$ of size $(\ngc - \sgc)$, an optimal decoding vector is given by~\eqref{eq:vopt-BIBD-code} and the worst-case approximation error for $\sgc$ stragglers is given by~\eqref{eq:errs-BIBD-code}.
\end{theorem}
\begin{IEEEproof}
The proof of Theorem~\ref{thm:BIBD-code} relies on the property that any two blocks of a symmetric BIBD intersect in exactly $\lambda$ points. Since the same property holds for the dual of a BIBD, the proof is identical to that of Theorem~\ref{thm:BIBD-code}. 
\end{IEEEproof}

Note that codes constructed from duals of BIBDs also admit computationally efficient decoding and are resilient to adversarial straggling by the same arguments as in Remark~\ref{rem:symmetric-BIBD-code-adversarial}.

It is possible to construct several classes of gradient codes by considering duals of well-known families of BIBDs. We present a few examples in the following.

\subsubsection{Class of Gradient Codes Based on the Duals of Affine Geometries} For any power of prime $q$ and integer $m\geq 2$, the affine geometry of order $q$ and dimension $m$ can be used to obtain a BIBD~\cite{Stinson:2003}. Using the dual of such a BIBD in Construction~\ref{con:BIBD-code} yields a class of gradient codes with the following parameters: $\ngc = q^m$, $\kgc = q(q^{m}-1)/(q-1)$, $\lgc = (q^{m}-1)/(q-1)$, $\rgc = q^{m-1}$ such that any pair of distinct workers share $\lamd = (q^{m-1}-1)/(q-1)$ gradient computations. 

\subsubsection{Class of Gradient Codes Based on the Duals of Residual and Derived Designs} Derived and residual BIBDs are well-known methods to obtain new BIBDs from symmetric BIBDs (see~\cite[Chapter 2.2]{Stinson:2003}). Using the duals of these designs allows us to construct gradient codes for a broad class of parameters. 

\subsection{Gradient Codes Using Resolvable Designs}
\label{sec:code-r-bibd}
Consider a gradient code with replication factor $\rgc$. If the number of stragglers $\sgc < \rgc$, then every gradient is computed by at least one of the remaining workers. Note that any exact gradient code can recover the gradient sum in this case. Therefore, it is desirable to construct approximate gradient codes that can exactly recover the gradient sum whenever $\sgc < \rgc$. However, as we can see from~\eqref{eq:errs-BIBD-code} that this is not the case for gradient codes obtained using either symmetric BIBDs or dual designs.

In this section, we consider gradient codes based on a special class of block designs called resolvable designs that lift this limitation. We begin with the definition of a resolvable BIBD.

\begin{definition}
\label{def:resolvable-bibd}
[{\it Resolvable BIBD}] A {\it parallel class} in a design is a subset of disjoint blocks  whose union is the point set. Let $(X,\mathcal{A})$ be a $\bibd$-BIBD. A partition of $\mathcal{A}$ into $\rd$ parallel classes is called a {\it resolution}. 
A $\bibd$-BIBD is said to be a {\it resolvable} BIBD if $\mathcal{A}$ has at least one resolution.
\end{definition}

\begin{remark}
\label{rem:r-bibd}
Note that a parallel class contains exactly $\vd/\kd$ blocks. Further, for any resolvable BIBD, it must be that $\bd\geq \vd+\rd-1$, or, equivalently, $\rd \geq \kd + \lambda$ (known as Bose's inequality).
\end{remark}

In the rest of this section, we focus our attention to a well-studied class of resolvable BIBDs called {\it affine resolvable BIBDs}.
A resolvable $\bibd$-BIBD with $\bd = \vd + \rd - 1$ (or, equivalently, $\rd = \kd + \lamd$) is said to be an {\it affine resolvable} BIBID.

\begin{example}
\label{ex:affine}
[{\it Affine Plane of Order 2}] A resolvable $(9,12,3,4,1)$-BIBD $X = \{1,2,\ldots,9\}$ and $\mathcal{A} = \{\mathcal{P}_1,\mathcal{P}_2,\mathcal{P}_3,\mathcal{P}_4\}$, where $ \mathcal{P}_1 = \{123,456,789\}$, $\mathcal{P}_2=\{147,258,369\}$, $\mathcal{P}_3=\{159,267,348\}$, and $\mathcal{P}_4=\{168,249,357\}$. 
Note that each $\set{P}_i$ is a parallel class and the partition $\{\set{P}_1,\ldots,\set{P}_4\}$ forms a resolution.
\end{example}


Consider a gradient code obtained from a resolvable $\bibd$-BIBD using Construction~\ref{con:BIBD-code}.
The resulting code is an $(\ngc = \bd, \kgc = \vd, \lgc = \kd, \rgc = \rd)$-GC. 
\begin{remark}
\label{rem:r-BIBD-disjoint}
Given an arbitrary resolution of the blocks $\set{A}$ as $\{\set{P}_1,\set{P}_2,\ldots,\set{P}_{\rd}\}$, the $\ngc$ workers can be partitioned into $\rgc$ $(= \rd)$ sets $\{\set{T}_1,\set{T}_2,\ldots,\set{T}_{\rgc}\}$ such that the $j$-th worker is included in set $\set{T}_i$ if the $j$-th block is in the set $\set{P}_i$. Then, naturally, for every $i\in[\rgc]$, any pair of distinct workers in $\set{T}_i$ compute disjoint gradients. Moreover, for every part $\set{T}_i$, $i\in[\rgc]$, workers in $\set{T}_i$ together compute all the gradients. Therefore, if the server receives the results from all the workers in any $\set{T}_i$, it can exactly recover the gradient sum by simply adding the partial gradient sums.
\end{remark}



\begin{remark}
\label{rem:r-bibd}
Consider an affine resolvable $\bibd$-BIBD $(X,\set{A})$ with incidence matrix $\mat{M}$. It is well-known that any two blocks from different parallel classes of an affine resolvable BIBD intersect in exactly $\kd^2/\vd$ points.\footnote{The parameters of an affine BIBD are such that $\kd^2/\vd$ is an integer.}
Therefore, any worker from $\set{T}_i$ shares exactly $\lgc^2/\kgc$ gradients with any worker from $\set{T}_j$ such that $j\ne i$. For simplicity, define $\mu: = \lgc^2/\kgc$.
\end{remark} 



The above property enables us to characterize the approximation error as well as optimal decoding vector in closed form. However, the analysis in this case turns out to be more intricate than the case of symmetric (or dual) BIBDs.\footnote{The increased complexity of analysis can be attributed to the fact that any pair of workers for an affine design share either zero gradients or $\lgc^2/\kgc$ gradients. On the other hand, for a symmetric (or dual) BIBD, any pair of workers share the same number of gradients. In fact, affine resolvable designs belong to a class of designs called {\it quasi-symmetric} designs: designs with the property that any pair of blocks intersect in either $x$ or $y$ points.} Towards this, we need to introduce the following notation.

\begin{definition}
\label{def:straggler-profile}
Consider a set of non-stragglers $\set{F}$ of size $(\ngc - \sgc)$. Define $\set{F}_i := \set{F}\cap\set{T}_i$ 
and $\sgc_i := \ngc/\lgc - |\set{F}_i|$. Note that $\sgc_i$ denotes the number of stragglers among the workers from $\set{T}_i$, 
and that $0\leq \sgc_i \leq \ngc/\lgc$ and $\sum_{i=1}^{\rgc}\sgc_i = \sgc$.  We call $[\sgc_1 \:\: \sgc_2 \:\: \cdots \:\: \sgc_{\rgc}]$ as the {\it straggler profile} corresponding to the set $\set{F}$. 
\end{definition}

\begin{theorem}
\label{thm:r-BIBD-code}
Consider an $(\ngc,\kgc,\lgc,\rgc)$-GC with encoding matrix $\mat{E}$ obtained from an affine resolvable BIBD using Construction~\ref{con:BIBD-code}. Consider a set of non-stragglers $\set{F}$ of size $(\ngc - \sgc)$ with straggler profile $[\sgc_1 \:\: \sgc_2 \:\: \cdots \:\: \sgc_{\rgc}]$. Define $\hat{\sgc}_0 := 0$ and $\hat{\sgc}_i := \sum_{p=1}^{i} \left(\kgc/\lgc - \sgc_p\right)$ for $i\in[\rgc]$. Recall that $\mu := \lgc^2/\kgc$.
\begin{enumerate}
    \item  If there exists an $i\in[\rgc]$ such that $\sgc_i = 0$, then an optimal decoding vector is 
\begin{equation}
    \label{eq:vopt-r-BIBD-code-1}
    \vopt(j) = 
    \begin{cases}
    1 & \textrm{for}\:\: \hat{\sgc}_{i-1}+1\leq j\leq \hat{\sgc}_{i}\\
    0 & \textrm{otherwise},
    \end{cases}
\end{equation}
    and the corresponding approximation error is $\errF{\mat{E}} = 0$.
    \item  Suppose $\sgc_i > 0 $ for all $i\in[\rgc]$, then an optimal decoding vector is 
\begin{equation}
    \label{eq:vopt-r-BIBD-code-2}
    \vopt(j) = \frac{\lgc/\left(\lgc - \mu(\kgc/\lgc - \sgc_i)\right)}{\left(1 + \sum_{p=1}^{\rgc}\frac{\mu(\kgc/\lgc - \sgc_p)}{\lgc - \mu(\kgc/\lgc - \sgc_p)}\right)} 
\end{equation}
    for $\hat{\sgc}_{i-1}+1\leq j\leq \hat{\sgc}_{i}$, and the corresponding approximation error is 
    \begin{IEEEeqnarray}{rCl}
    \label{eq:errF-r-BIBD-code}
    \errF{\mat{E}} & = & \kgc + 2\sum_{i = 1}^{\rgc} \lgc(\kgc/\lgc - \sgc_i)c_i(c_i-2)\nonumber\\
    & {} &\: +\sum_{i=1}^{\rgc}\sum_{\substack{j = 1\\ j\ne i}}^{\rgc}\mu(\kgc/\lgc - \sgc_i)(\kgc/\lgc - \sgc_j)c_i c_j\IEEEeqnarraynumspace,
    \end{IEEEeqnarray}
    where $c_i = \frac{\lgc/\left(\lgc - \mu(\kgc/\lgc - \sgc_i)\right)}{\left(1 + \sum_{p=1}^{\rgc}\frac{\mu(\kgc/\lgc - \sgc_p)}{\lgc - \mu(\kgc/\lgc - \sgc_p)}\right)}$.
\end{enumerate} 
\end{theorem}
\begin{IEEEproof}
(Sketch) For the first part, 
note that all the workers from $\set{T}_i$ return their computations. From Remark~\ref{rem:r-BIBD-disjoint}, 
simply summing the results from the workers in $\set{T}_i$ recovers the gradient sum. The vector in~\eqref{eq:vopt-r-BIBD-code-1} computes this sum. Clearly, the approximation error in this case is zero.

For the second part, we show that it is possible to solve the normal equation by computing $\mat{E}^{\dagger}$ in closed formed. For this, we leverage the block intersection property of affine resolvable designs mentioned in Remark~\ref{rem:r-bibd}. In this case, we need to iteratively use the {\it matrix inversion lemma} given in Appendix~\ref{app:matrix-inversion-lemma}. The complete proof is deferred to Appendix~\ref{app:r-BIBD-code}. 
\end{IEEEproof}

\begin{remark}
\label{rem:r-BIBD-code-adversarial}
{
Note that since optimal decoding vector depends only on the straggler profile (number of stragglers from each set $\set{T}_i$, $i\in[\rgc]$), decoding at the server can be performed in a very efficient way. Further, any adversary that can enforce at most $\sgc_i$ stragglers from set $\set{T}_i$, $i\in[\rgc]$, cannot worsen the error by  intelligently selecting stragglers as opposed to randomly selecting stragglers.
}
\end{remark}


\begin{table}[!ht]
  \renewcommand{\arraystretch}{1.2}
  \caption{Gradient Codes Using Block Designs. The parameter $q$ is a power of a prime and $m \geq 2$ is an integer.}
  \label{tab:table_example}
  \centering
  \begin{tabular}{|c|c|c|c|}
    \hline
    Class & Design & Parameters of GC\\
    \hline
    \hline
    \begin{tabular}{@{}c@{}} Symmetric\\ 
     BIBD \end{tabular} & 
     \begin{tabular}{@{}c@{}} Projective Geometry\\ 
     (PG) \end{tabular} & 
     \begin{tabular}{@{}c@{}} $\ngc = \kgc = \frac{(q^{m+1} - 1)}{(q-1)}$, \\ 
     $\lgc = \rgc = \frac{(q^m - 1)}{(q - 1)}$
    \end{tabular} \\
    \hline
    \begin{tabular}{@{}c@{}} Dual of a\\ 
     BIBD \end{tabular} & 
     \begin{tabular}{@{}c@{}} Dual of Affine Geometry\\ 
     (Dual AG) \end{tabular} & 
    \begin{tabular}{@{}c@{}} $\ngc = q^m$, $\kgc = q\frac{(q^{m}-1)}{(q-1)}$, \\ 
     $\lgc = \frac{(q^{m}-1)}{(q-1)}$, $\rgc = q^{m-1}$
    \end{tabular} \\
    \hline
    \begin{tabular}{@{}c@{}} Resolvable\\ 
     BIBD \end{tabular} & 
     \begin{tabular}{@{}c@{}} Affine Geometry\\ 
     (AG) \end{tabular} & 
    \begin{tabular}{@{}c@{}} $\ngc = q\frac{(q^{m}-1)}{(q-1)}$, $\kgc = q^m$, \\ 
     $\lgc = q^{m-1}$, $\rgc = \frac{(q^{m}-1)}{(q-1)}$
    \end{tabular} \\
    \hline
  \end{tabular}
\end{table}


\subsection{Summary of Constructions}
\label{sec:summary-constructions}

It is possible to construct several classes of gradient codes based on well-known families of BIBDs. We summarize a few examples in Table~\ref{tab:table_example}. 

\begin{figure*}[!ht]
     \subfloat[]{%
       \includegraphics[scale=0.28]{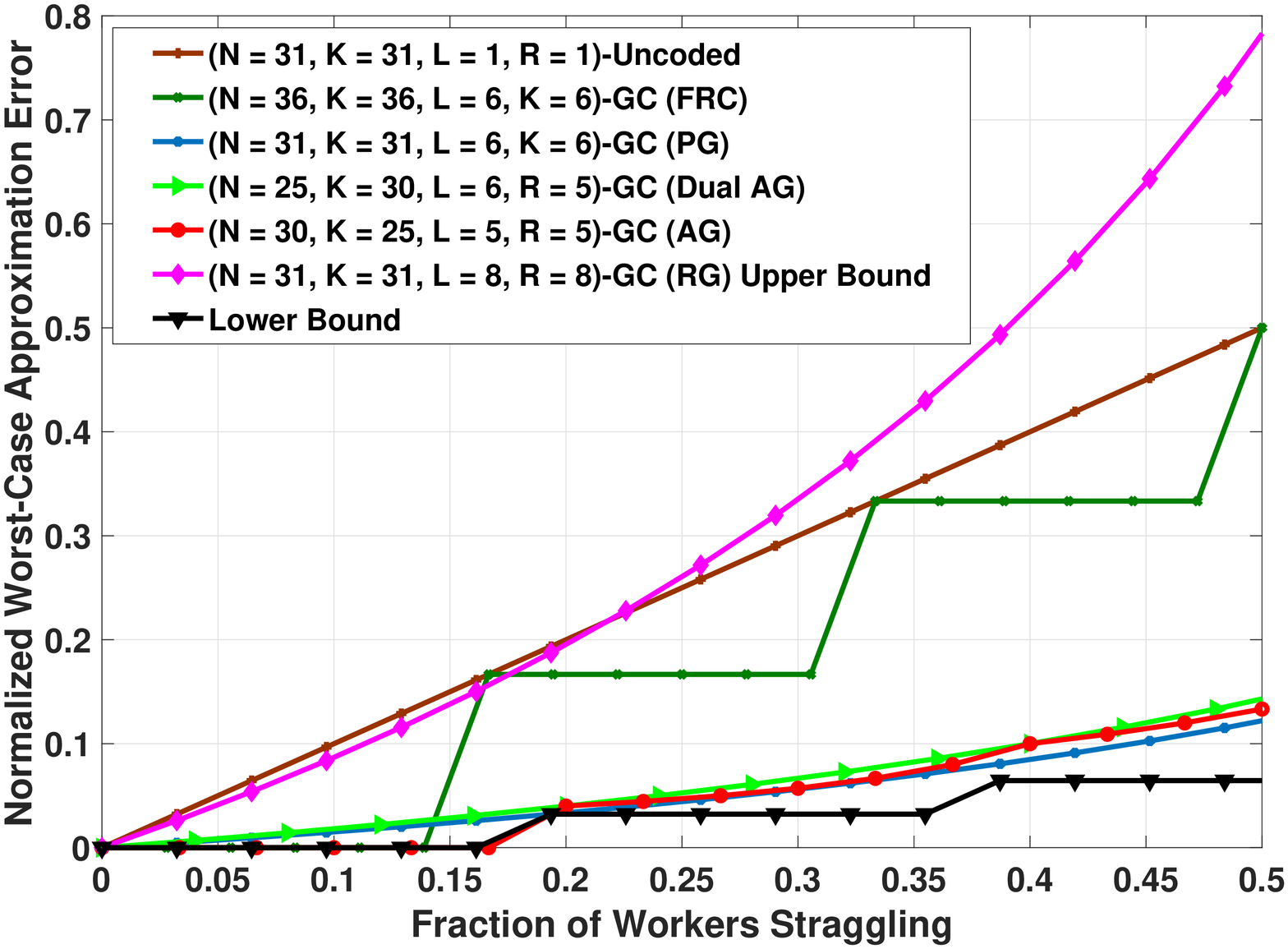} 
     }\hfil
     \subfloat[]{%
       \includegraphics[scale=0.28]{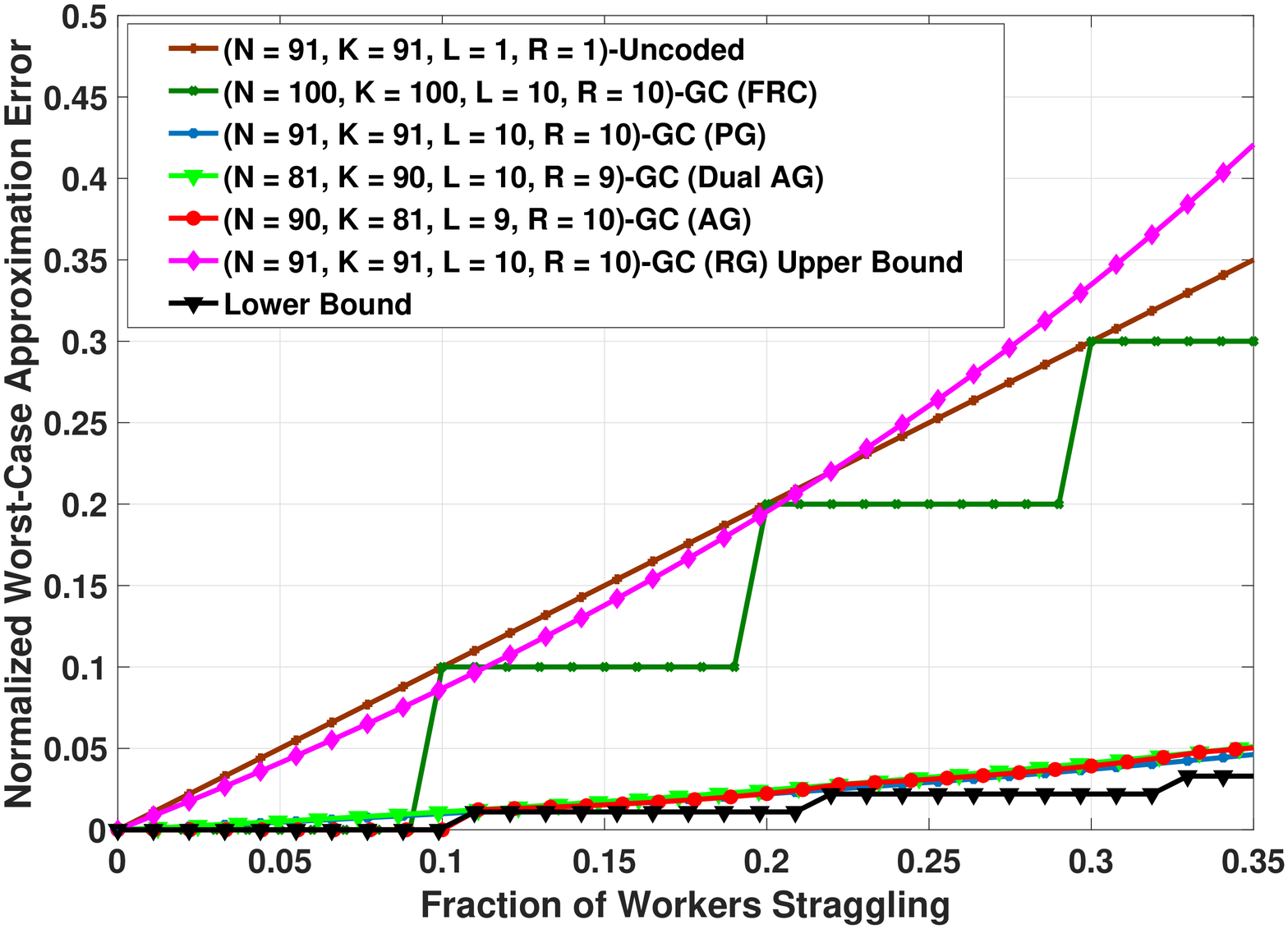}
     }
     \caption{Performance evaluation of gradient coding schemes based on block designs.}
     \label{fig:numerical}
\end{figure*}

\section{Performance Evaluation}
\label{sec:numerical}

In this section, we numerically evaluate the performance of the proposed design-based schemes. 
We consider the following gradient coding schemes (see Table~\ref{tab:table_example}, fix $m = 2$): (i) $(\ngc = q^2+q+1, \kgc = q^2+q+1, \lgc = q+1, \rgc = q+1)$-GC based on the projective plane of order $q$ (denoted as PG), (ii) $(\ngc = q^2, \kgc = q^2+q, \lgc = q+1, \rgc = q)$-GC based on the dual of affine plane of order $q$ (denoted as Dual AG), and (iii) $(\ngc = q^2+1, \ngc = q^2, \lgc = q, \rgc = q+1)$-GC based on the affine plane of order $q$ (denoted as AG).  

We plot the worst-case approximation error normalized by the number of gradients, \ie, $\errs{\mat{E}}/\kgc$ versus the normalized number of stragglers, \ie, $\sgc/\ngc$. Specifically, we consider the following two regimes: $q = 5$ and $q = 9$ in Figures~\ref{fig:numerical}(a) and~\ref{fig:numerical}(b), respectively. Observe that the three families PG, Dual AG, and AG have similar performance in terms of approximation error.

For comparison, we plot the uncoded case which partitions $\kgc = q^2+q+1$ gradients across $\ngc = q^2+q+1$ workers. Note that the approximation error in this case equals the number of stragglers. 
In addition, we also consider an \mbox{$(\ngc = (q+1)^2, \kgc = (q+1)^2, \lgc = q+1, \rgc = q+1)$}-fractional repetition code (FRC)~\cite{Tandon:17,CharlesP:17}. As expected, both the uncoded and FRC schemes perform poorly when the stragglers are adversarial.

In addition, we consider codes based on Margulis construction of Ramanujan graphs in~\cite[Example 19]{Raviv:18}, denoted as RG. For these codes, we plot the upper bound on the worst-case approximation error derived in~\cite{Raviv:18} as a proxy for the worst-case approximation error. This is because, to obtain the worst-case approximation error, one needs to consider all possible subsets of stragglers. This becomes computationally infeasible for large $\ngc$ and $\sgc$. We see that the worst-case approximation error for BIBD-based codes is substantially smaller than the guarantees given by the upper bound for the RG scheme.

To see how well the proposed codes perform, we consider a lower bound on the worst-case approximation error from~\cite{Raviv:18}. In particular, in~\cite[Lemma 21]{Raviv:18}, the authors showed that for any $(\ngc,\kgc,\lgc,\rgc)$-GC $\mat{E}$ with $\kgc = \ngc$, the worst-case approximation error can be lower bounded as $\errs{\mat{E}}\geq \lfloor \sgc/\lgc \rfloor$. 
From Fig.~\ref{fig:numerical}(a) and~\ref{fig:numerical}(b), we can observe that the proposed schemes perform close to this lower bound for the small number of stragglers. {It is worth noting that, in massive-scale serverless systems, which are our motivation to mitigate adversarial stragglers, only a small number of machines straggle substantially (see, \eg,~\cite[Fig.~1]{Gupta:oversketch:18}).}

In fact, gradient codes based on projective planes are nearly optimal for large $q$ and $\sgc = O(q)$.
To see this, consider \mbox{$(\ngc = q^2+q+1, \kgc = q^2+q+1, \lgc = q+1, \rgc = q+1)$-GC} based on the projective plane of order $q$. The worst-case approximation error in~\eqref{eq:errs-BIBD-code} reduces to the following expression.
\begin{equation}
    \label{eq:errs-PG-2}
    \errs{\mat{E}_{\textrm{PG}}} = \frac{\sgc}{(q+1)+\frac{q+1-\sgc}{q}}
\end{equation}
Observe that when $S = O(q)$ and $q$ is large, the error above is close to the lower bound $\lfloor \sgc/(q+1) \rfloor$.

\section{Robustness Against Adversarial Straggling}
\label{sec:adversarial}
Our goal in this section is to investigate fundamental limits on the approximation error. As mentioned in the previous section, for any $(\ngc,\kgc,\lgc,\rgc)$-GC $\mat{E}$ with $\kgc = \ngc$, the worst-case approximation error can be lower bounded as $\errs{\mat{E}}\geq \lfloor \sgc/\lgc \rfloor$~\cite[Lemma 21]{Raviv:18}. In fact, the proof of~\cite[Lemma 21]{Raviv:18} is constructive and gives an $O(\ngc^2)$ time greedy algorithm to find a set of stragglers that will enforce $\errs{\mat{E}}\geq \lfloor \sgc/\lgc \rfloor$. In other words, even an adversary with limited computing power can induce the error of at least $\lfloor \sgc/\lgc \rfloor$. However, in general, for a given gradient code and a number $S$, finding a set of $S$ stragglers that maximize the approximation error is shown to be NP-hard in~\cite{CharlesP:17}.\footnote{The authors consider the case when the decoding vector is fixed {\it a priori}. In particular, it is assumed that the decoding vector is of the form $\mat{v} = \rho\Jc{\ngc-\sgc}$ for a fixed constant $\rho$.}



To analyze fundamental limits 
for a computationally unbounded adversary, 
we consider the following problem: given a gradient code and a {\it target} $\eta$, what is the minimum number of stragglers that an adversary must introduce to ensure that the approximation error is at least $\eta$? 


Towards this, consider a bipartite graph $\set{G} = (\set{W},\set{D},\set{E})$ for a given  $(\ngc,\kgc,\lgc,\rgc)$-GC with encoding matrix $\mat{E}$ as follows. The left $\ngc$ vertices $\set{W}$ correspond to the set of workers, while the right $\kgc$ vertices $\set{D}$ correspond to the set of gradients to be computed. There is an edge $\{i,j\}\in\set{E}$ from a vertex $i\in\set{W}$ to a vertex $j\in\set{D}$ iff $\mat{E}_{i,j} \ne 0$. Note that the graph $\set{G}$ specifies the {\it placement scheme} for the gradient code, \ie, how the data parts are assigned to the workers.

Consider a set $\set{T}\subset{\set{D}}$ and let $\Neb{\set{T}}\subset\set{W}$ denote the neighbors of $\set{T}$ in $\set{G}$. Now, suppose all the workers in $\Neb{\set{T}}$ are straggling. Then, the gradients in $\set{T}$ cannot contribute to the gradient sum. Therefore, the approximation error must be at least $|\set{T}|$. Based on this observation, we introduce the notion of {\it adversarial threshold} by defining the following {\it adversarial straggling problem}. 

\begin{definition}
\label{def:adversarial-threshold}
[{\it Adversarial Threshold}]
Given a graph $\set{G}$ associated with an $(\ngc,\kgc,\lgc,\rgc)$-GC and a constant $0 < \eta < \kgc$, define
\begin{equation}
    \label{eq:S-star}
     \sgt{\eta} : = \arg\min_{\substack{\set{T}\subset \set{D}\\ |\set{T}| = \eta}} |\Neb{\set{T}}|.
\end{equation}
We refer to refer to the above minimization problem as the {\it adversarial straggling problem}, and $\sgt{\eta}$ as the {\it adversarial threshold}.
\end{definition}

Note that, given $\set{G}$, $\sgt{\eta}$ is the smallest number of workers that must be selected by an adversarial straggler to enforce that the approximation error is at least $\eta$.\footnote{We do not consider an encoding matrix $\mat{E}$ explicitly in the formulation for simplicity.}

Next, we derive a lower bound on $\sgt{\eta}$.
We restrict our attention on a class $\code$  of gradient codes for which $\ngc = \kgc$, and the associated bipartite graph $\set{G}$ is regular and connected. 

\begin{proposition}
\label{prop:S-star-LB}
For any gradient code from the class $\code$, and for any $\eta \leq \ngc/4$, we have
\begin{equation}
\label{eq:S-star-LB}
\sgt{\eta} \geq \left(\frac{3\lgc - \lamd_2}{\lgc + \lamd_2}\right) \eta =: \sgc^{*}_{LB}(\eta),
\end{equation}
where $\lamd_2$ is the second largest eigenvalue of the graph $\set{G}$ associated with the code.\footnote{For a brief review of eignevalues of a graph, see Appendix~\ref{app:graphs}.}
\end{proposition}
\begin{IEEEproof}
See Appendix~\ref{app:S-star-LB}.
\end{IEEEproof}

Next, we show that codes obtained from symmetric BIBDs are 
{excellent candidates to mitigate adversarial stragglers}, since they achieve the maximum $\sgt{\eta}$ among the codes from $\code$.

\begin{proposition}
\label{prop:BIBD-optimal}
Let $\eta \leq \ngc/4$. 
Gradient codes obtained from symmetric BIBDs via Construction~\ref{con:BIBD-code} achieve the maximum value of $\sgt{\eta}$ among the codes in $\set{C}$.
\end{proposition}
\begin{IEEEproof}
See Appendix~\ref{app:BIBD-optimal}.
\end{IEEEproof}





\bibliographystyle{IEEEtran}
\bibliography{Grad_coding_designs}

\appendices
\section{Matrix Inversion Lemma}
\label{app:matrix-inversion-lemma}
\begin{lemma}
\label{lem:matrix-inversion}  
(cf.~\cite{Miller:81:matrix-inversion}) Let $\mat{G}$ and $\mat{G}+\mat{H}$ be nonsingular matrices where $\mat{H}$ is a matrix of rank one. Then, $\mathrm{tr}\left(\mat{H}\mat{G}^{-1}\right)\ne -1$, and the inverse of $(\mat{G}+\mat{H})$ is
\begin{equation}
    \label{eq:matrix-inversion}
    (\mat{G}+\mat{H})^{-1} = \mat{G}^{-1} - \frac{1}{1+\mathrm{tr}\left(\mat{H}\mat{G}^{-1}\right)}\mat{G}^{-1}\mat{H}\mat{G}^{-1}.
\end{equation}
\end{lemma}

\section{Proof of Theorem~\ref{thm:BIBD-code}}
\label{app:BIBD-code}
Consider an arbitrary set of non-stragglers $\set{F}\subset[N]$ of size $(\ngc - \sgc)$. Define $\nsgc := \ngc - \sgc$. Recall that we have
\begin{equation*}
    \label{eq:vopt}
    \vopt = \arg\min_{\mat{v}\in\mathbb{R}^{\ngc-\sgc}}\norm{\mat{E}_{\set{F}}\mat{v} - \Jc{\kgc}}.
\end{equation*}
One optimal solution to the above least squares problem is $\vopt = \mat{E}_{\set{F}}^{\dagger}\Jc{\kgc}$.

Since $\mat{E} = \mat{M}$, each column of $\mat{E}$ contains exactly $\lgc$ ones and any two columns of $\mat{E}$ intersect in exactly $\lamd$ locations (see Remark~\ref{rem:incidence-matrix}). Therefore, we have 
\begin{IEEEeqnarray}{rCl}
\label{eq:E-times-1}
\matEF^T\Jc{\kgc} & = & \lgc\Jc{\nsgc},\\
\label{eq:E-times-E}
\matEF^T\matEF & = & 
(\lgc - \lamd)\I{\nsgc} + \lamd\Js{\nsgc}.
\end{IEEEeqnarray}

Note that the matrix on the right hand side of~\eqref{eq:E-times-E} above has an eigenvalue $\lgc - \lamd$ with multiplicity $\nsgc - 1$ and an eigenvalue $(\lgc - \lamd) + \lamd\nsgc$ with multiplicity one. Thus, its determinant is $(\lgc - \lamd)^{\nsgc-1}((\lgc - \lamd) + \lamd\nsgc) \ne 0$.\footnote{For any BIBD, $\kd > \lamd$. Therefore, we have $\lgc > \lamd$. Note that the same proof also works for Theorem~\ref{thm:dual-BIBD-code}, where we use a dual of a BIBD. In this case, since $\rd > \lamd$, we again have $\lgc > \lamd$.} Therefore, $\matEF^T\matEF$ is nonsingular, and we have $\matEF^{\dagger} = (\matEF^T\matEF)^{-1}\matEF^T$. 

Next, we compute $(\matEF^T\matEF)^{-1}$ as follows:
\begin{IEEEeqnarray}{rCl}
(\matEF^T\matEF)^{-1} & \stackrel{(a)}{=} & \left((\lgc - \lamd)\I{\nsgc} + \lamd\Js{\nsgc}\right)^{-1},\nonumber\\
& \stackrel{(b)}{=} & \frac{1}{\lgc - \lamd}\I{\nsgc} - \frac{1}{1 + \mathrm{tr}\left(\frac{\lamd}{\lgc - \lamd}\Js{\nsgc}\right)}\frac{\lamd}{(\lgc - \lamd)^2}\Js{\nsgc},\nonumber\\
& {=} & \frac{1}{\lgc - \lamd}\I{\nsgc} - \frac{1}{1 + \left(\frac{\lamd\nsgc}{\lgc - \lamd}\right)}\frac{\lamd}{(\lgc - \lamd)^2}\Js{\nsgc},\nonumber\\
\label{eq:EE-inverse}
& {=} & \frac{1}{\lgc - \lamd}\left[\I{\nsgc} - \frac{\lamd}{\lgc + \lamd(\nsgc-1) }\Js{\nsgc}\right],
\end{IEEEeqnarray}
where (a) follows from~\eqref{eq:E-times-E}, and (b) follows from Lemma~\ref{lem:matrix-inversion} in Appendix~\ref{app:matrix-inversion-lemma}. 

Now, we can compute $\vopt$ as
\begin{IEEEeqnarray}{rCl}
\vopt & \stackrel{(c)}{=} & (\matEF^T\matEF)^{-1}\matEF^T\Jc{\kgc},\nonumber\\
& \stackrel{(d)}{=} & \frac{1}{\lgc - \lamd}\left[\I{\nsgc} - \frac{\lamd}{\lgc + \lamd(\nsgc-1)}\Js{\nsgc}\right]\lgc\Jc{\nsgc},\nonumber\\
\label{eq:vopt-BIBD-code-2}
& {=} & \frac{\lgc}{\lgc + \lamd(\nsgc-1)}\Jc{\nsgc},
\end{IEEEeqnarray}
where (c) follows from $\vopt = \matEF^{\dagger}\Jc{\kgc}$, and (d) follows from~\eqref{eq:E-times-1}. 
Finally,~\eqref{eq:vopt-BIBD-code} follows from~\eqref{eq:vopt-BIBD-code-2} noting that $\nsgc = \ngc - \sgc$.

Next, we compute $\errF{\mat{E}}$ for an arbitrary set of non-stragglers $\set{F}$ of size $\nsgc$.
\begin{IEEEeqnarray}{rCl}
\errF{\mat{E}} 
& \stackrel{(e)}{=} & \left(\matEF\vopt - \Jc{\kgc}\right)^T\left(\matEF\vopt - \Jc{\kgc}\right),\nonumber\\
& {=} & \Jc{\kgc}^T\Jc{\kgc} - 2\vopt^T\matEF^T\Jc{K} + \vopt^T\matEF^T\matEF\vopt,\nonumber\\
& \stackrel{(f)}{=} & \kgc - 2\lgc\vopt^T\Jc{\nsgc} +   \vopt^T((\lgc - \lamd)\I{\nsgc} + \lamd\Js{\nsgc})\vopt,\nonumber\\
& \stackrel{(g)}{=} & \kgc - \frac{2\lgc^2\nsgc}{\lgc+\lamd(\nsgc-1)} +   \frac{\lgc^2((\lgc-\lamd)\nsgc+\lamd\nsgc^2)}{(\lgc+\lamd(\nsgc-1))^2},\nonumber\\
\label{eq:err-BIBD-code-2}
& {=} & K - \frac{\lgc^2\nsgc}{\lgc+\lamd(\nsgc-1)},
\end{IEEEeqnarray}
where (e) follows from $\errF{\mat{E}} = \norm{\matEF\vopt - \Jc{\kgc}}$, (f) follows from~\eqref{eq:E-times-1} and~\eqref{eq:E-times-E}, and (g) follows after substituting $\vopt$ from~\eqref{eq:vopt-BIBD-code-2}.

Since $\errF{\mat{E}}$ does not depend on the specific set of stragglers, but only the size of it, we get~
\eqref{eq:errs-BIBD-code} from~\eqref{eq:err-BIBD-code-2} substituting $\nsgc = \ngc - \sgc$.

\section{Proof of Theorem~\ref{thm:r-BIBD-code}}
\label{app:r-BIBD-code}
Consider an arbitrary set of non-stragglers $\set{F}$ of size $(\ngc - \sgc)$ with straggler profile $[\sgc_1 \:\: \sgc_2 \:\: \cdots \:\: \sgc_{\rgc}]$. Recall that $0\leq \sgc_i \leq \ngc/\lgc$ and $\sum_{i=1}^{\rgc}\sgc_i = \sgc$. Define $\nsgc_i := \kgc/\lgc - \sgc_i$ for $i\in[\rgc]$ and $\nsgc_0 := 0$. We consider the second case when $\sgc_i > 0$ for every $i\in[\rgc]$. 

Recall that we need to solve
\begin{equation*}
    \label{eq:vopt}
    \vopt = \arg\min_{\mat{v}\in\mathbb{R}^{\ngc-\sgc}}\norm{\mat{E}_{\set{F}}\mat{v} - \Jc{\kgc}}.
\end{equation*}
One optimal solution to the above least squares problem is $\vopt = \mat{E}_{\set{F}}^{\dagger}\Jc{\kgc}$.

By following the proof of Bose's inequality for resolvable block designs, we have that any sub-matrix of $\mat{E}$ with an arbitrary column removed from each of $\set{T}_1,\set{T}_2,\ldots,\set{T}_{\rgc}$ has full column rank. Since we have $\sgc_i > 0$ for every $i\in[\rgc]$, it follows that $\matEF$ has full column rank. Therefore, $\matEF^T\matEF$ is nonsingular, and we have $\matEF^{\dagger} = (\matEF^T\matEF)^{-1}\matEF^T$. 

From Remark~\ref{rem:r-bibd}, we obtain that
\begin{IEEEeqnarray}{rCl}
\label{eq:r-E-times-1}
\matEF^T\Jc{\kgc} & = & \lgc\Jc{\nsgc},\\
\label{eq:r-E-times-E}
\matEF^T\matEF & = & \Jh + \mu\Js{\nsgc}, 
\end{IEEEeqnarray}
where $\Jh$ is a block matrix defined as
\begin{equation}
    \label{eq:J-hat}
    \Jh = 
    \begin{bmatrix}
    \Jhs{\nsgc_1} & {} & {} & {}\\
    {} & \Jhs{\nsgc_2} & {} & {}\\
    {} & {} & {\ddots} & {}\\
    {} & {} & {} & \Jhs{\nsgc_{\rgc}}
    \end{bmatrix}
\end{equation}
such that $\Jhs{\nsgc_i} = \lgc\I{\nsgc_i} - \mu\Js{\nsgc_i}$ for $i\in[\rgc]$.
Note that we suppress the zero entries in the right hand side of~\eqref{eq:J-hat} for simplicity.

Next, from~\eqref{eq:r-E-times-E} and Lemma~\ref{lem:matrix-inversion} (in Appendix~\ref{app:matrix-inversion-lemma}), we get:
\begin{equation}
\label{eq:r-EE-inverse-1}
(\matEF^T\matEF)^{-1} = {\Jh}^{-1} - \frac{1}{1 + \mathrm{tr}\left(\mu\Js{\nsgc}{\Jh}^{-1}\right)}{\Jh}^{-1}\mu\Js{\nsgc}{\Jh}^{-1}.
\end{equation}
Due to the block structure of $\Jh$, we have
\begin{equation}
    \label{eq:J-hat-inv}
    \Jh^{-1} = 
    \begin{bmatrix}
    \Jhs{\nsgc_1}^{-1} & {} & {} & {}\\
    {} & \Jhs{\nsgc_2}^{-1} & {} & {}\\
    {} & {} & {\ddots} & {}\\
    {} & {} & {} & \Jhs{\nsgc_{\rgc}}^{-1}
    \end{bmatrix},
\end{equation}
where $\Jhs{\nsgc_i}^{-1}$ can be computed as follows:
\begin{IEEEeqnarray}{rCl}
\Jhs{\nsgc_i}^{-1} 
& = & \left(\lgc\I{\nsgc_i} + (-\mu)\Js{\nsgc_i}\right)^{-1},\nonumber\\
& \stackrel{(a)}{=} & \frac{1}{\lgc}\I{\nsgc_i} - \frac{1}{1+\textrm{tr}\left(-\mu\Js{\nsgc_i}\frac{1}{\lgc}\I{\nsgc_i} \right)}\frac{1}{\lgc}\I{\nsgc_i}(-\mu)\Js{\nsgc_i}\frac{1}{\lgc}\I{\nsgc_i},\nonumber\\
\label{eq:Jhs-inv}
& = & \frac{1}{\lgc}\I{\nsgc_i} + \frac{\mu}{\lgc(\lgc - \mu\nsgc_i)}\Js{\nsgc_i}.
\end{IEEEeqnarray}
The equality (a) above  is obtained using Lemma~\ref{lem:matrix-inversion}. 

Using~\eqref{eq:J-hat-inv} and~\eqref{eq:Jhs-inv}, we verify that
\begin{equation}
    \label{eq:J-Jh-inv}
    \mu\Js{\nsgc}\Jh^{-1} = 
    \begin{bmatrix}
    \left(\frac{\mu}{\lgc - \mu\nsgc_1}\right)\J{\nsgc}{\nsgc_1} & 
    \cdots &
    \left(\frac{\mu}{\lgc - \mu\nsgc_{\rgc}}\right)\J{\nsgc}{\nsgc_{\rgc}}
    \end{bmatrix},
\end{equation}
and thus, $\textrm{tr}\left(\mu\Js{\nsgc}\Jh^{-1}\right) = \sum_{p=1}^{\rgc}\frac{\mu\nsgc_p}{\lgc - \mu\nsgc_p}$. Using~\eqref{eq:J-Jh-inv}, one can verify that
\begin{equation}
\label{eq:Jh-inv-J-Jh-inv}
    \Jh^{-1}\mu\Js{\nsgc}\Jh^{-1} =
    \begin{bmatrix}
    \mat{A}_{1,1} &  \mat{A}_{1,2} & \cdots & \mat{A}_{1,\rgc}\\
    \mat{A}_{2,1} &  \mat{A}_{2,2} & \cdots & \mat{A}_{2,\rgc}\\
    \mat{A}_{\rgc,1} &  \mat{A}_{\rgc,2} & \cdots & \mat{A}_{\rgc,\rgc}
    \end{bmatrix},    
\end{equation}
where 
\begin{equation}
    \label{eq:Aij}
    \mat{A}_{i,j} = \mu\left(\frac{1}{\lgc - \mu\nsgc_i}\right)\left(\frac{1}{\lgc - \mu\nsgc_j} \right)\J{\nsgc_i}{\nsgc_j}.
\end{equation}

We verify, by substituting the above results  in~\eqref{eq:r-EE-inverse-1} and using~\eqref{eq:r-E-times-1}, that $(\matEF^T\matEF)^{-1}\matEF^T\Jc{\kgc}$ results in the expression of $\vopt$ given in~\eqref{eq:vopt-r-BIBD-code-2}.

Finally, note that 
\begin{IEEEeqnarray}{rCl}
\errF{\mat{E}} 
& {=} & \left(\matEF\vopt - \Jc{\kgc}\right)^T\left(\matEF\vopt - \Jc{\kgc}\right)\nonumber\\
\label{eq:errF-r-BIBD-code-0}
& {=} & \Jc{\kgc}^T\Jc{\kgc} - 2\vopt^T\matEF^T\Jc{K} + \vopt^T\matEF^T\matEF\vopt.
\end{IEEEeqnarray}
Expression~\eqref{eq:errF-r-BIBD-code} is obtained by using~\eqref{eq:r-E-times-1}, \eqref{eq:r-E-times-E}, and~\eqref{eq:vopt-r-BIBD-code-2} in~\eqref{eq:errF-r-BIBD-code-0}.

\section{Overview of Expansion and Spectral Properties of a Graph}
\label{app:graphs}
Let $\set{G} = (\set{V},\set{E})$ be a finite, undirected and connected graph on $\ngc$ vertices. For a subset of vertices $\set{F}\subset\set{V}$, the {\it boundary} $\partial\set{F}$ is the set of edges connecting $\set{F}$ to $\set{V}\setminus\set{F}$. The {\it expanding constant} or {\it isopetimetric constant} of $\set{G}$ is defined as (see~\cite{Hoory06expandergraphs})
\begin{equation}
    \label{eq:isoperimetric-constant}
    h(\set{G}) = \min_{\emptyset\ne\set{F}\subset\set{V}}\frac{|\partial\set{F}|}{\min\{|\set{F}|,|\set{V}\setminus\set{F}|\}}.
\end{equation}

It is well-known that expansion properties of a graph are closely related to the {\it adjacency matrix} $\mat{A}$ of the graph, defined as follows:  
$\mat{A}_{i,j} = 1$ iff vertices $i$ and $j$ are connected by an edge, \ie, $\{i,j\}\in\set{E}$. Since $\mat{A}$ is a real symmetric matrix, it has real eigenvalues $\lamd_1 \geq \lamd_2 \geq \cdots \geq \lamd_{\ngc}$. When $\set{G}$ is $\lgc$-regular, it is well-known that $\lamd_1 = \lgc$ , $\lamd_2 < \lgc$  and $\lamd_{\ngc} \geq -\lgc$, where equality holds iff $\set{G}$ is bipartite~\cite{Hoory06expandergraphs}. 

\begin{theorem}
\label{thm:expansion}
(cf.~\cite{Alon:86})
If $\set{G}$ is a finite, connected, $\lgc$-regular graph, then
\begin{equation}
    \label{eq:expansion}
    \frac{\lgc - \lamd_2}{2} \leq h(\set{G}) \leq \sqrt{2\lgc(\lgc - \lamd_2)}.
\end{equation}
\end{theorem}

Connected regular graphs for which $\lamd_2$ is smaller than the vertex degree are called as {\it expanders}. 

\begin{theorem}
\label{thm:lamd-2-lower-bound}
(cf.~\cite{Hoholdt:12}) Let $\set{G} = (\set{V},\set{E})$ be a connected, $(\lgc,\rgc)$-regular graph. Then
\begin{equation}
    \label{eq:lamd-2-lower-bound}
    \lamd_2 \geq \left(\frac{|\set{E}| - \lgc\rgc}{\frac{|\set{E}|}{\lgc} - 1} \right)^{\frac{1}{2}}.
\end{equation}
For the $\rd$-regular graph of a symmetric $\bibd$-BIBD, the bound in~\eqref{eq:lamd-2-lower-bound} is satisfied with equality.
\end{theorem}

\section{Proof of Proposition~\ref{prop:S-star-LB}}
\label{app:S-star-LB}
Let $\set{G} = (\set{W},\set{D},\set{E})$ be the associated bipartite graph. Consider any $\set{T}\subset\set{D}$ of size $\eta$ and let $\set{S} = \Neb{\set{T}}\subset\set{W}$. Let $S = |\set{S}|$. Note that there are $\eta\lgc$ edges from $\set{T}$ to $\set{S}$. Further, there are $S\lgc$ edges such that, for each edge, one of the endpoints is incident on $\set{S}$. Let $\partial(\set{S}\cup\set{T})$ be the boundary of $\set{S}\cup\set{T}$. Recall that this is the set of edges connecting $\set{S}\cup\set{T}$ to $\{\set{W}\cup\set{D}\}\setminus\{\set{S}\cup\set{T}\}$. Therefore, we have $\lgc S = \lgc \eta + |\partial(\set{S}\cup\set{T})|$, from which, we get
\begin{IEEEeqnarray}{rCl}
S & = & \eta + \frac{1}{\lgc}|\partial(\set{S}\cup\set{T})|\\
& = & \eta + \left(\frac{S+\eta}{\lgc}\right)\left(\frac{|\partial(\set{S}\cup\set{T})|}{S+\eta}\right)\\
& \stackrel{(a)}{\geq} & \eta + \frac{S+\eta}{\lgc}h(\set{G})\\
\label{eq:S-star-LB-2}
& \stackrel{(b)}{\geq} & \eta + \left(\frac{S+\eta}{\lgc}\right)\left(\frac{\lgc - \lamd_2}{2}\right),
\end{IEEEeqnarray}
where (a) follows from~\eqref{eq:isoperimetric-constant} and $\eta\leq\ngc/4$, and (b) follows from Theorem~\ref{thm:expansion}. The result follows after rearranging~\eqref{eq:S-star-LB-2}.

\section{Proof of Proposition~\ref{prop:BIBD-optimal}}
\label{app:BIBD-optimal}
From~\eqref{eq:S-star-LB}, observe that the smaller the $\lamd_2$ the larger is $\sgtLB{\eta}$. Therefore, the result follows immediately from Theorem~\ref{thm:lamd-2-lower-bound}.


\IEEEtriggeratref{3}


\end{document}